\documentclass{article}
\usepackage{graphicx}
\usepackage{amsmath}
\usepackage{amssymb}
\usepackage{subfigure}
\usepackage{url}
\usepackage{wrapfig}
\usepackage{hyperref}

\newcommand{\hide}[1]{{}}
\def \modediagram {{\textsf{MDM}}}
\def \State{\textit{State}}
\def \MD {\textit{md}}
\def \mlist{\textit{mlist}}
\def \TopModes {\emph{TopModes}}
\def \Contains {\textit{Contains}}
\def \Trans {\textit{Trans}}

\def \PBegin   {{\textit{Begin}}}
\def \PExecute {{\textit{Execute}}}
\def \PEnd     {{\textit{End}}}
\def \CFGStart {{\textit{Start}}}
\def \CFGExit  {{\textit{Exit}}}
\def \sampling {{\mathsf{sampling}}}
\def \execute  {{\textit{execute}}}
\def \failure  {{\textit{failure}}}

\newcommand{\submode}[2]{\mathsf{sub\_mode}(#1, #2)}
\newcommand{\supermodes}[2]{\mathsf{super\_modes}(#1, #2)}
\newcommand{\upmodes}[3]{\mathsf{up\_modes}(#1, #2, #3)}
\newcommand{\outs}[1]{\mathsf{outs}(#1)}
\newcommand{\CHOP} {{^{\frown}}}

\def \kwVar {\textit{Var}}
\def \kwMode {\textit{Mode}}
\def \kwModule {\textit{Module}}
\def \kwname {\textit{name}}
\def \kwperiod {\textit{period}}
\def \kwinitial {\textit{initial}}
\def \kwBody {\textit{Body}}
\def \kwTransition {\textit{Transition}}
\def \kwguard {\textit{guard}}
\def \kwpriority {\textit{priority}}
\def \kwtarget {\textit{target}}
\def \kwsource {\textit{source}}
\def \kwCFG {\textit{CFG}}

\def \kwSystem {\textit{md}}
\def \kwConst {\textit{Const}}
\def \kwSExpr {\textit{SExpr}}
\def \kwBTerm {\textit{BTerm}}
\def \kwBExpr {\textit{BExpr}}
\def \kwIExpr {\textit{IExpr}}
\def \kwGTerm {\textit{GTerm}}
\def \kwGuard {\textit{guard}}
\def\sigmasq{\ensuremath{\sigma_1{\cdot}...{\cdot}\sigma_n}}

\def \SA {\emph{A}}
\def \SB {\emph{B}}

\newcommand{\period}[1]{\mathsf{period(#1)}}
\newcommand{\isInitial}[1]{\mathsf{is\_initial(#1)}}
\newcommand{\CFG}[1]{\mathsf{CFG(#1)}}

\newcommand{\priority}[1]{\mathsf{prio(#1)}}
\newcommand{\guard}[1]{\mathsf{guard(#1)}}
\newcommand{\target}[1]{\mathsf{target(#1)}}
\newcommand{\source}[1]{\mathsf{source(#1)}}

\newcommand{\True}{\mathsf{\mathsf{true}}}
\newcommand{\False}{\mathsf{\mathsf{false}}}
\newcommand{\tvalue}{\textit{tt}}
\newcommand{\fvalue}{\textit{ff}}
\newcommand{\timestamp}{\textit{timestamp}}

\newcommand{\Int}{\mathcal{I}nt}

\newcommand{\Terms}{\textit{Terms}}
\newcommand{\Formulas}{\textit{Formulas}}
\newcommand{\Intv}{\textit{Intv}}

\newcommand{\after}{\mathsf{after}}
\newcommand{\duration}{\mathsf{duration}}

\newcommand{\runa}[1]{\mbox{\textsc{\protect{(#1})}}}

\newcommand{\infrulesL}[3]
           {\parbox{#1mm}{$$ \frac{#2}{#3}\hspace{.5cm}$$}}

\def \stmts    {\textit{stmts}}
\def \aStmt    {\textit{aStmt}}
\def \pStmt    {\textit{pStmt}}
\def \cStmt    {\textit{cStmt}}
\newcommand{\CALL}{\mathsf{call}}
\newcommand{\WHILE}{\mathsf{while}}
\newcommand{\DO}{\mathsf{do}}
\newcommand{\IF}{\mathsf{if}}
\newcommand{\THEN}{\mathsf{then}}
\newcommand{\ELSE}{\mathsf{else}}

\newcommand{\SKIP}{\mathsf{skip}}

\newtheorem{property}{\textbf{Property}}

\newcommand{\autref}[1]{\if\thepage0\else\hyperlink{aut#1}{$^#1$}\fi}
\newcommand{\sponsor}[1]{\if\thepage0\else\footnote{#1}\fi}

\newcommand{\email}[1]{\href{mailto:#1}{#1}}

\title
{
    {\modediagram}: A Mode Diagram Modeling Framework for \\ Periodic Control Systems
}

\author
{
    Zheng Wang$^{1, 5}$,
    Geguang Pu$^{1}$,
    Shenchao Qin$^{2}$,
    Jianwen Li$^{1}$, \\
    Kim G. Larsen$^{3}$,
    Jan Madsen$^{4}$,
    Bin Gu$^{5}$,
    Jifeng He$^{1}$
}

\begin{document}

\maketitle

\begin{center}
$^{1}$ \email{wangzheng@sei.ecnu.edu.cn}, \email{ggpu@sei.ecnu.edu.cn},\\
             Shanghai Key Laboratory of Trustworthy Computing,\\
             East China Normal University \\
$^{2}$ \email{s.qin@tees.ac.uk},  University of Teesside \\
$^{3}$ \email{kgl@cs.aau.dk},  Aalborg University of Denmark \\
$^{4}$ \email{jan@imm.dtu.dk},  Technical University of Denmark \\
$^{5}$ \email{gubin88@yahoo.com.cn},  Beijing Institute of Control Engineering
\end{center}

\begin{abstract}

Periodic control systems  used in spacecrafts and automotives are usually period-driven and can be decomposed into different modes with each mode representing a system state observed from outside. Such  systems may also involve  intensive computing in their modes.
Despite the fact that such control systems are widely used in the above-mentioned safety-critical embedded  domains, there is lack of domain-specific formal modelling languages for such systems in the relevant industry.
To address this problem, we propose a formal visual modeling framework called  {\modediagram} as a concise and precise way to specify and analyze such systems. To capture the temporal properties of  periodic control systems, we provide, along with {\modediagram},  a property specification language based on interval logic for the description of concrete temporal requirements the engineers are concerned with. The statistical model checking technique can then be used to verify the {\modediagram} models against desired properties. To demonstrate the viability of our approach, we have applied our modelling framework to some real life case studies from industry and helped detect two design defects for some spacecraft control systems.
\end{abstract}


\section{Introduction}

\hide{Periodic control systems are widely used in embedded computing area, for instance, spacecraft control system and automotive control system etc. Such systems are usually driven by a global or local clock, which triggers the behavior of the system in periods. Another feature of these systems is that they can be decomposed into several different modes from the system view.  Each mode actually represents an important state of the system. Further more, a mode can also be divided into several sub-modes. When a period ends, the system may switch from the current mode to another mode if the transformation condition holds. For instance, China Academy of Space Technology (CAST) spends great efforts in designing and developing embedded software for spacecrafts. The periodic control system is a key for the spacecrafts, which has the following features:}%

Control systems that are widely used in safety-critical embedded domains, such as spacecraft control and automotive control, usually reveal periodic behaviours. Such {\em periodic} control systems share some interesting features and characteristics:
\begin{figure*}[t]
    \centering
    \includegraphics[width=0.98\textwidth]{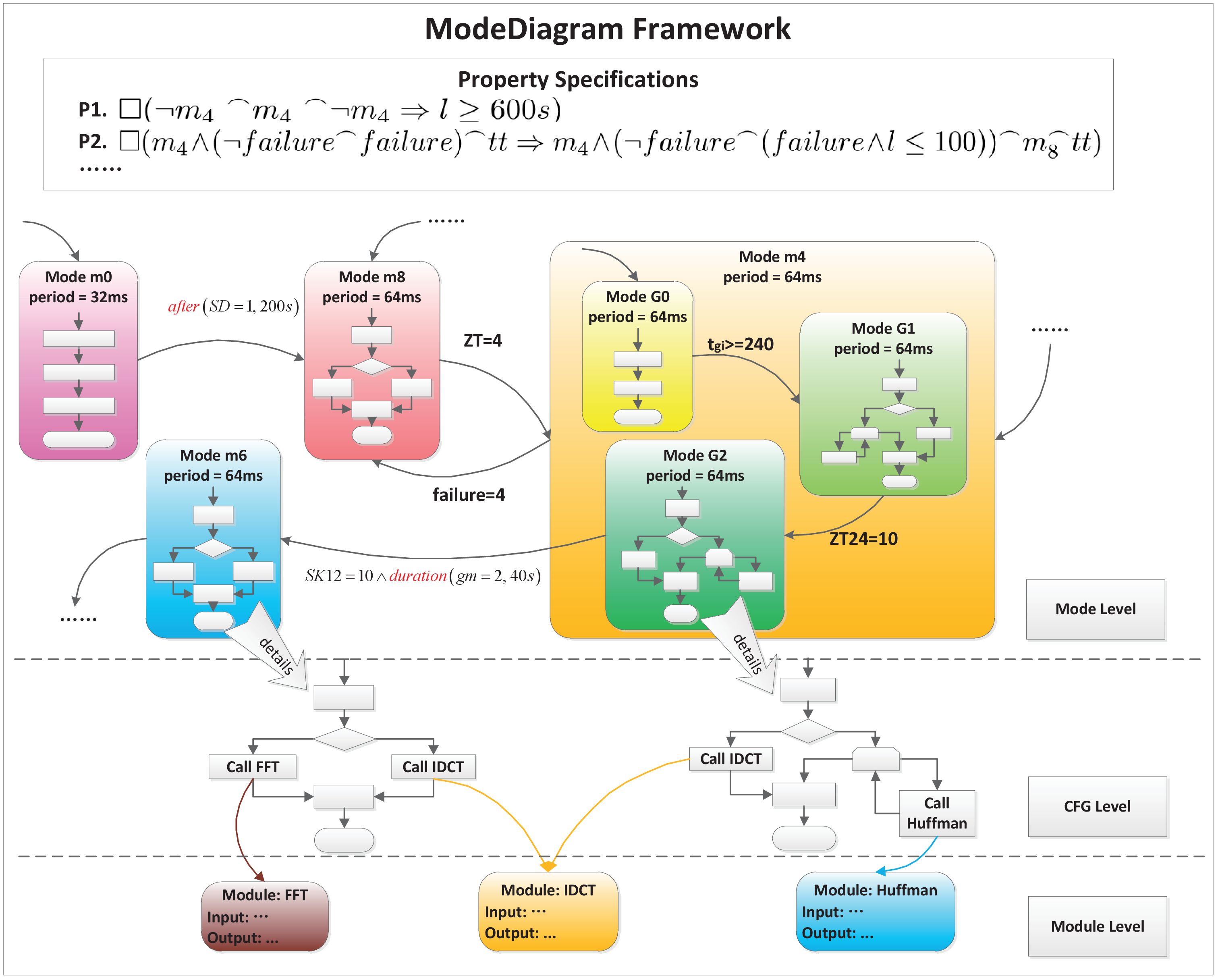}\\
    \vspace*{-3mm}
    \caption{ {\modediagram}: An (Incomplete) Example }\label{fig:mcoverall}\vspace*{-3mm}
\end{figure*}

\begin{itemize}
    \item They are {\em mode-based}.  A periodic control system is usually composed of a set of modes, with each mode representing an important state of the system. Each mode either contains a set of sub-modes or performs controlled computation periodically.
    \item They are {\em computation-oriented}. In each mode,  a periodic control system may perform control algorithms involving complex computation. For instance, in certain mode, a spacecraft control system may need to process intensive data in order to decide its space location.

    \item They behave {\em periodically}.  A periodic control system is reactive and may run for a long time. The behaviour of each mode is regulated by its own period. That is, most computations are performed within a  period and may be repeated in the next period if mode switch does not take place. Mode switch may only take place at the end of a period under certain conditions.

\end{itemize}

Despite the fact that periodic control systems have been widely used in areas such as spacecraft control, there is lack of a concise and precise domain specific formal modelling language for such systems. In our joint project with China Academy of Space Technology (CAST), we have started with several existing modelling languages but they are either too complicated therefore require too big a learning curve for domain engineers, or  are too specific/general, therefore require non-trivial restrictions or extensions.  This motivates us to propose  a new formal but lightweight modelling language that matches exactly the need of the domain engineers,  the so called  \underline{M}ode \underline{D}iagram \underline{M}odelling framework ({\modediagram}).

Although the proposed modelling notation {\modediagram} can be regarded as a variant of Statecharts \cite{statechart}, it has been specifically designed to cater for the domain-specific need in modelling periodic control systems. We shall now use an example  to illustrate informally the {\modediagram} framework, and leave the formal syntax and semantics to the next section. As shown in Fig~\ref{fig:mcoverall}, the key part of a {\modediagram} model is the collection of modes given in the mode level.  Each mode has a period, and the periods for different modes can be different. A mode can be nested and the transitions between modes or sub-modes may take place. A transition may be enabled  if the associated guard is satisfied. In {\modediagram}, the transition guards may involve complex temporal expressions. For example, in the transition from mode {\sf G2} to mode {\sf m6}, in addition to the condition $\textsf{SK12=10}$, it also requires that the condition $\textsf{gm=2}$ has held for 40s, as captured by the $\mathsf{duration}$ predicate.

An {\modediagram} model is presented hierarchically. A mode that does not contain any sub-modes (termed a {\em leaf} mode) contains a control flow graph (CFG) encapsulating specific control algorithms or computation tasks. The details of CFGs are given in the CFG level. The CFGs may refer to modules (similar to procedures in conventional languages) details of which are given in the module level.

To support formal reasoning about {\modediagram} models, we also provide a property specification language inspired by an interval-like calculus~\cite{IntervalCalculus}, which facilitates the capture of temporal properties  system engineers may be interested in. Two example properties are listed in Fig~\ref{fig:mcoverall}. The property {\sf P1} says that ``whenever the system enters the {\sf m4} mode, it should stay there for at least 600s''. The formal details of the specification language is left to a later section.

To reason about whether an {\modediagram} model satisfies  desired properties specified by system engineers using the property specification language, we employ statistical model checking techniques~\cite{SMCYounes05,SMCYounesS02}. Since {\modediagram} may involve complex non-linear computation in its flow graph, the complete verification is undecidable. Apart from incompleteness, statistical model checking can verify  hybrid systems efficiently \cite{StatHeterogeneous}. Our experimental results on real life cases have demonstrated that the statistical model checking can help uncover  potential defects of {\modediagram} models.

In summary, we have made the following contributions in this paper:

\begin{itemize}\setlength{\itemsep}{0pt}
    \item  We propose a novel visual formal modelling notation {\modediagram}    as a concise yet precise modelling language for  periodic control systems.
    \item  We present a formal semantics for {\modediagram} and a property specification language to facilitate the verification process.
    \item We develop a new statistical model checking algorithm   to verify   {\modediagram} models against various temporal properties.
    \item  We have carried out real life case studies  to demonstrate the effectiveness of the proposed framework.  We have managed to discover some  design defects of a real spacecraft control  system.
\end{itemize}

The rest of this paper is organized as follows. Section~\ref{sec:modechart} presents the formal syntax and semantics of {\modediagram}. Section~\ref{sec:specification} introduces our interval-based property specification language and its semantics. The statistical model checking algorithm for the {\modediagram} is developed in Section~\ref{sec:statistical}. Implementations and case studies are presented in Section~\ref{sec:implementation}, followed by related work and concluding remarks. \hide{We also discuss related work in Section~\ref{sec:discussion}. Finally, we conclude this paper in Section~\ref{sec:conclusion}. }

\hide{
To specify periodic control systems precisely, we introduce a new modeling notation named ``{\modediagram}'', which can be considered as a variance of StateChart\cite{statechart}. It provides the basic modeling element \emph{mode} that represents the observable state in the control system. Each mode has a period, and the periods of different modes can be different. The mode can be nested and the transitions among modes or sub-modes can take place. The condition of the transition may involve complex temporal expressions. Fig.~\ref{fig:mcoverall} shows an example of {\modediagram}. In this figure, before the mode $G_2$ switches to mode $m_6$, the formula $SK12 = 10$ should be satisfied and the formula $gm = 2$ has to hold in the last 400 seconds, which is captured by the predicate $\mathsf{duration}$. The structure of {\modediagram} is hierarchical and the top level is the modes (sub-modes). In leaf mode, it may have the flow graph which encapsulates control algorithms or other computation tasks. In flow graph, it provides the procedure unit, which is similar to the function concept in programming language for the encapsulation. {\modediagram} provides a visual notation for the engineers to specify the control system conveniently and precisely. Since the period is an important issue for the control system, we take it as a first-class element in {\modediagram}. That is, all the computations or transitions in a mode of {\modediagram} are driven by the given period.

{\modediagram} is actually a domain-specific visual language~\cite{Liu2005visual} for embedded systems. The idea behind the {\modediagram} is to combine the state-oriented tasks and computation-oriented ones in a period-driven framework.  Furthermore, {\modediagram} also provides a property specification language to facilitate to capture the temporal properties the system engineers are interested in. For instance, a property like ``\emph{When the system switch to mode $M_2$, it should stay in this mode at least 600ms}''. When we study this property, we can find the the property involves the duration concept \emph{``at least 600ms"}.

So  we adopt an interval-like calculus~\cite{IntervalCalculus} as the property specification language, which involves temporal variables and intervals. For instance,  The following interval calculus formula captures the property above precisely and Fig.~\ref{fig:intervalexample} shows the property looks like in the time line.
\vspace*{-2mm}
\[
\Box(\neg m_4\ ^{\frown} m_4\ ^{\frown} \neg m_4 \Rightarrow l \geq
600s)
\]

For a given interval, this formula requires that for any sub-interval, if the system is not in $m_4$ at first, after entering mode $m_4$ the system will leave $m_4$ eventually, which  implies that the length of this sub-interval should be at least 600 seconds.

\begin{figure}[t]
    \centering
    \includegraphics[scale=0.9]{pic/intervalexample.eps}\\
    \caption{An Interval Property}\label{fig:intervalexample}
\end{figure}

To verify whether the {\modediagram} satisfies the system properties specified by its property specification language, we apply statistical model checking technique~\cite{SMCYounesS02,SMCYounes05} to obtain this goal. Since {\modediagram} may involve complex non-linear computation in its flow graph, the complete verification is undecidable. However, statistical model checking can verify the hybrid system efficiently though it has its shortage on the complete verification. The experimental results on real life cases show that the statistical model checking can uncover the potential defaults of the control system.

Although {\modediagram} is inspired from the periodic control systems used in spacecrafts, it is extensible to other domains, such as automotive electronics. In this paper, our work makes the four contributions:

\begin{itemize}
    \item  A novel notation {\modediagram}  with its property specification language is proposed for the description of periodic control systems.
    \item The formal semantics of {\modediagram}  is presented to clarify its precise behaviors and also facilitate the verification process
    \item A new statistical model checking algorithm based on {\modediagram}  is developed to verify the temporal properties the control system should obey.
    \item The real life cases are carried out to show the effectiveness of the proposed approach.  In the empirical experiments,  the design defects of the real
     system have been revealed by our approach.
\end{itemize}

The rest of this paper is organized as follows. Section~\ref{sec:modechart} gives the formal syntax and semantics to {\modediagram}. Section~\ref{sec:specification} introduces an interval-oriented specification language, and interprets its semantics on the traces model. The statistical model checking algorithm for the {\modediagram} is developed in Section~\ref{sec:statistical}. Implementations and case studies are presented in Section~\ref{sec:implementation}. We also discuss related work in Section~\ref{sec:discussion}. Finally, we conclude this paper in Section~\ref{sec:conclusion}.

}

\section{The {\modediagram} Notations}\label{sec:modechart}

\subsection{The Syntax of {\modediagram}}

\begin{figure}[t]
    \centering
    \begin{tabular}{c}
    $
    \begin{small}
        \begin{array}{rcl}
          \kwSystem     & ::= & (\kwVar^{+}, \kwMode^{+}, \kwModule^{+})  \\
          \kwMode       & ::= & (\kwname, \kwperiod, \kwinitial, \kwBody, \kwTransition^{+})  \\
          \kwBody       & ::= & \kwMode^{+} \mid \kwCFG \\
          \kwTransition & ::= & (\kwsource,\kwguard, \kwpriority, \kwtarget) \\
          \kwModule     & ::= & (\kwname, V_I, V_O, CFG)
        \end{array}
    \end{small}
    $ \\
    (a){\modediagram} \\
    \\
    $
    \begin{small}
        \begin{array}{rcl}
            \kwSExpr & ::= & \kwConst \mid \kwVar \mid f^{(n)}(\kwSExpr \ldots) \\
            \kwBTerm & ::= & \True \mid \False \mid p^{(n)}(\kwSExpr \ldots) \\
            \kwIExpr & ::= & (\after \mid \duration) ( \kwBTerm, \kwSExpr ) \\
            \kwGTerm & ::= & \kwIExpr \mid \kwBTerm \\
            \kwBExpr & ::= & \hspace*{1mm} \kwBTerm \mid \neg \kwBExpr \mid \kwBExpr \vee \kwBExpr \mid \kwBExpr \wedge \kwBExpr \\
            \kwGuard & ::= & \hspace*{1mm} \kwGTerm \mid \neg \kwGuard \mid \kwGuard \vee \kwGuard\mid \kwGuard \wedge \kwGuard \\
        \end{array}
    \end{small}
    $
    \\
    (b) Expressions and Guards \\
    \\
    $
            \begin{array}{rl}
                 CFG =_{df} &  \stmts \\
              \stmts =_{df} &  \pStmt \mid \cStmt\\
              \pStmt =_{df} &  \aStmt \mid \CALL ~ name \mid \SKIP ~ |\\
              \aStmt =_{df} &  x := \kwSExpr \\
              \cStmt =_{df} &  \stmts ;~\stmts \mid  \WHILE ~ \kwBExpr ~ \DO ~ \stmts \mid  \\
                            &  \IF ~ \kwBExpr ~  \THEN ~ \stmts ~ \ELSE ~ \stmts
            \end{array}
        $
    \\
    (c) CFG \\
    \end{tabular}
    \caption{The Syntax of {\modediagram}}\label{fig:mcsyntax}\vspace*{-5mm}
\end{figure}

We briefly list its syntactical elements in Fig.~\ref{fig:mcsyntax}(a). An {\modediagram} is composed of a list of modes ($\kwMode^+$) and modules ($\kwModule^{+}$), as well as a list of variables ($\kwVar^+$) used in those modes and modules.

Intuitively, a mode  refers to a certain state of the system which can be observed from outside. A mode has a name, a period, a body and a list of transitions. For simplicity, we assume all mode names are distinct in an MDM model. The mode period (a real number) is used to trigger the periodic behavior of the  mode. The mode body can be composed of either a control flow graph (CFG), prescribing the computational tasks the system can perform in the mode in every period, or a list of other modes as the immediate sub-modes of the current mode. If a mode has sub-modes, when the control lies in  this mode, the control should also be in one of the sub-modes. A leaf mode does not have sub-modes, so  its body contains a CFG. A mode is either a leaf mode, or it directly or indirectly has  leaf modes as its sub-modes. A mode is called top mode if it is not a sub-mode of any other mode. The CFG in a leaf mode is the standard control flow graph, which contains nodes and structures like assignment, module call, conditional and loop. It also supports  function units like the ones in conventional programming languages. The syntax of CFG is presented in Figure~\ref{fig:mcsyntax}(c).

A module encapsulates computational tasks as its CFG. $V_I$ specifies the set of variables used in the CFG, while $V_O$ is the set of variables modified in the CFG. A module can be invoked by some modes or other modules. As a specification for embedded systems, recursive module calls are forbidden.

A transition (from $\kwTransition^+$) specifying a mode switch from  one mode to another is represented as a quadruple, where the first element is the name of the source mode, the second specifies the transition condition,  the third  is the priority of the transition and the last element is the name of the target mode. The {\modediagram} supports mode switches at different levels in the mode-hierarchy. The transition condition (i.e. $\kwguard$) is defined in Fig.~\ref{fig:mcsyntax}(b). A state expression can be either a constant, a variable, or a real-value function on state expressions. A boolean term is either a boolean constant, or a predicate on state expressions. There are two kinds of interval expressions, $\mathsf{after}$ and $\mathsf{duration}$. These interval expression are very convenient to model system behaviors related with past states. A guard term can be either an interval expression, or a boolean term. A guard is the boolean combination of guard terms. To ensure the mode switches deterministic, we require that the priority of a transition should be distinguish with others in the same mode chain:
\[
    \forall m \in \kwMode \cdot \forall t_1, t_2 \in \outs{\supermodes{\MD}{m}} \cdot
    t_1 \ne t_2 \Rightarrow \priority{t_1} \ne \priority{t_2}
\]
The functions $\supermodes{\MD}{m}$ and $\outs{\mlist}$ will be defined later.

\subsubsection{Auxiliary Definitions}

Given an {\modediagram}
$
\begin{small}
\kwSystem ::= (\kwVar^{+}, \kwMode^{+}, \kwModule^{+})
\end{small}
$,
we introduce two auxiliary relations:\\
$Contains(\kwSystem) \subseteq {\kwMode}s \times {\kwMode}s$ for mode-subsume relation and\\
$Trans(\kwSystem) \subseteq {\kwMode}s \times \Int \times \kwGuard \times {\kwMode}s$ for mode-switch  relation.

Given a mode $m = (n, per, ini, b, tran)$ and a transition $t = (m, g, pri, m')$, we define these operations/predicates:
\[
    \begin{array}{llll}
        \period{m} = per   ~&~ \isInitial{m} = ini ~& ~\CFG{m} = b   ~&~                 \\
        \priority{t} = pri ~&~ \guard{t} = g       ~& ~\source{t}=m  ~&~ \target{t} = m'  \\
    \end{array}
\]
We also define the following auxiliary  functions:
\[
    \begin{array}{l}
        \supermodes{\MD}{m} \triangleq \langle m_1, m_2, \ldots, m_k \rangle, \text{ where }\\
        \hspace*{5mm} m_k = m \wedge m_1 \in {\TopModes}(\MD) \wedge \forall 1 {<} i {\le} k \cdot (m_{i-1}, m_i) \in \Contains(\MD) \\
        \hspace*{5mm} \text{and } m{\in}{\TopModes}(\MD) \triangleq  m {\in} {\kwMode}s(\MD) {\wedge} \neg\exists m'\hide{ {\in} {\kwMode}s}\cdot ( m', m ) { \in} \Contains(\MD) \\
        \\
       \upmodes{\MD}{m}{k} \triangleq \{ m_i \mid m_i \in \supermodes{\MD}{m} \wedge \mod(k, \frac{\period{\mathit{m_i}}}{\period{\mathit{m}}}) = 0 \} \\
       \\
       \submode{\MD}{m} \triangleq m', \text{ where } (m, m') \in \Contains \wedge \isInitial{\mathit{m'}} \\
       \\
       \outs{\mlist} \triangleq \bigcup_{m \in \mlist}\{t \mid t \in \Trans \wedge \source{\mathit{t}}=m\}
    \end{array}
\]
The function $\supermodes{\MD}{m}$  retrieves a sequence of modes from a top mode to $m$ using the $\Contains$ relation. The set ${\TopModes}(\MD)$ consists all the modes which are not sub-modes of any other mode. The function $\upmodes{\MD}{m}{k}$ returns those modes in $\supermodes{\MD}{m}$ whose periods are consistent with the period count $k$. {\modediagram} requires that the period of a mode should be equal to or multiple to the period of its sub-modes. The function $\submode{\MD}{m}$ returns the initial sub-mode for a non-leaf node $m$, and the predicate $\isInitial{\mathit{m'}}$ means that the sub-mode $m'$ is the initial sub-mode in its hierarchy. The function $\outs{\mlist}$ returns all outgoing transitions from modes in $\mlist$.

\subsection{The Semantics}\label{subsec:semantics}

In order to precisely analyze the behaviors of {\modediagram}, for instance, model checking of {\modediagram} , we need its formal semantics. In this section, we present the operational semantics for {\modediagram}.

\begin{table}[t]
    \[
    \begin{small}
    \begin{array}{lcl}
      \sigmasq \models b                          & \Leftrightarrow & \sigma_n \models b \\
      \sigmasq \models \neg g                     & \Leftrightarrow & \neg(\sigmasq \models g)  \\
      \sigmasq \models g_1 \vee g_2               & \Leftrightarrow & \sigmasq \models g_1 \text{ or } \sigmasq \models g_2 \\
      \sigmasq \models g_1 \wedge g_2             & \Leftrightarrow & \sigmasq \models g_1 \text{ and } \sigmasq \models g_2 \\
      \sigmasq \models \mathsf{duration}(b, l)    & \Leftrightarrow & \sigma_n(l) = \nu \wedge  \exists i {<}n \cdot ( \sigma_i(ts) {+} \nu\leq \sigma_n(ts) \wedge \\
                                                                  &                   &
       \sigma_{i{+}1}(ts) {+} \nu \geq \sigma_n(ts) \wedge       \forall i {\leq} j {\leq} n \cdot \sigma_j(b) = \True)  \\
      \sigmasq \models \mathsf{after}(b, l)       & \Leftrightarrow & \sigma_n(l) = \nu \wedge       \exists i {<} n \cdot (   \sigma_i(ts) {+}\nu \leq \sigma_n(ts) \wedge\\
                                                                  &                   &
 \sigma_{i{+}1}(ts) {+} \nu \geq \sigma_n(ts)) \wedge      \sigma_i(b) = \True) 
    \end{array}
    \end{small}
    \]
    \caption{The Interpretation of Guards}\vspace*{-6mm}
    \label{tbl:evalie}
\end{table}

\subsubsection{Configuration}

The configuration in our operational semantics is represented as $( \MD, m, l, pc, k, \Sigma )$, where
\begin{itemize}
    \item $\MD$ is the \modediagram, and $m$ is the  mode  the system control currently lies in.
    \item $l \in \{\PBegin, \PExecute, \PEnd\}$ specifies the system is in the beginning, middle, or end of a period.
    \item $pc \in \mathcal{L}$, where $\mathcal{L} = \mathcal{N} \cup \{\CFGStart, \CFGExit, \bot\}$ is the program counter to execute the control flow graph. $\mathcal{N}$ is used to represent the nodes in control flow graphs and $\CFGStart$, $\CFGExit$ denote the start and exit location of a control flow graph respectively. If the current mode is not equipped with any flow graph, we use the symbol $\bot$ as a placeholder.
    \item The fourth component $k$ records the count of periods for the current mode. If the system switches to another mode, it will be reset to $1$. The period count is used to distinguish whether a super-mode of the current mode is allowed to check its mode switch guard.
    \item $\Sigma$ is a list of states of the form $\Sigma' \cdot \sigma$, where $\sigma$ denotes the current state ($\sigma\in \State \triangleq \textit{Vars}{\rightarrow}\mathbb{R}$) and $\Sigma'$ represents a history of states.

\end{itemize}

\noindent{\bf Guards\quad}The evaluation of a transition guard may depend on the current state as well as some history states. Table~\ref{tbl:evalie} shows how to interpret a guard in a given sequence of states. The symbol $ts$ is the abbreviation of the variable $\timestamp$. The guard $\mathsf{duration}(b,l)$ evaluates to $\True$ if the boolean expression $b$ has been  $\True$  during the time interval $l$ up to the current moment. The guard  $\mathsf{after}(b,l)$ evaluates to $\True$ if  the boolean expression $b$ was $\True$  the time interval $l$ ago. In this table, $b$ is a pure boolean expression
without interval expressions and $l$ is a state expression.

\subsubsection{Operational Rules}
\label{subsec:inferencerules}

\begin{table}[t]
    \centering
    \[
    \begin{small}
    \begin{array}{cc}
        \runa{enter} & \infrulesL{120}
                        {
                            \CFG{m} = \bot
                        }
                        {
                            ( \MD, m, \PBegin, \bot, k, \Sigma )
                            \overset{}{\longrightarrow}
                            ( \MD, m', \PBegin, pc', k, \Sigma )
                        }\\
                        & \hspace*{-8mm}\text{where }  m' = \submode{\MD}{m} \text{ and }
                                         pc' =
                                            \begin{cases}
                                                \bot ,     & \mbox{if } \CFG{m'} = \bot \\
                                                \CFGStart, & \mbox{if } \CFG{m'} \neq \bot
                                            \end{cases}  \\
        \runa{detect} &\hspace*{-8mm} \infrulesL{120}
                      {
                        \CFG{m} \ne \bot
                      }
                      {
                        ( \MD, m, \PBegin, pc, k, \Sigma \cdot \sigma )
                        \overset{}{\longrightarrow}
                        ( \MD, m, \PExecute, pc, k, \Sigma \cdot \sampling(\sigma) )
                      }\\
        \runa{execute} & \infrulesL{120}
                      {
                        \execute (\CFG{m}, pc, \sigma, \period{m}) = ( pc', \sigma' )
                      }
                      {
                        ( \MD, m, \PExecute, pc, k, \Sigma {\cdot}\sigma)
                        \overset{}{\longrightarrow}
                        ( \MD, m, \PEnd, pc', k, \Sigma' )
                      }\\
                      &
                        \begin{array}{cl}
                          \text{where } & \Sigma' = \Sigma \cdot \sigma'[ts \mapsto \sigma(ts) + \period{m}] \\
                        \end{array}   \\
        \runa{continue} & \infrulesL{120}
                      {
                        pc \neq \CFGExit
                      }
                      {
                        ( \MD, m, \PEnd, pc, k, \Sigma )
                        \overset{}{\longrightarrow}
                        ( \MD, m, \PExecute, pc, k, \Sigma )
                      }\\
        \runa{repeat} & \infrulesL{120}
                      {
                        \forall t \in \outs{\upmodes{\MD}{m}{k}} \cdot \Sigma \not\models \guard{t}
                      }
                      {
                        ( \MD, m, \PEnd, \CFGExit, k, \Sigma )
                        \overset{}{\longrightarrow}
                        ( \MD, m, \PBegin, \CFGStart, k{+}1, \Sigma )
                      }\\
        \runa{switch} &\hspace*{-8mm} \infrulesL{120}
                      {
                        \begin{array}{l}
                            \exists t \in \outs{\upmodes{\MD}{m}{k}} \cdot \Sigma \models \guard{t} \wedge \\
                          {\forall} t' {\in} \outs{\upmodes{\MD}{m}{k}} {-} \{t\} \cdot (\Sigma\not\models\guard{t'} \vee \priority{t'} {<} \priority{t})
                        \end{array}
                      }
                      {
                        ( \MD, m, \PEnd, \CFGExit, k, \Sigma )
                        \overset{}{\longrightarrow}
                        ( \MD, m', \PBegin, pc', 1, \Sigma )
                      }\\
                      & \text{where } m' = \target{t} \text{ and } pc' =
                                            \begin{cases}
                                                \bot,      & \mbox{if } \CFG{m'} = \bot \\
                                                \CFGStart, & \mbox{if } \CFG{m'} \neq \bot
                                            \end{cases} \\
                      &
    \end{array}
    \end{small}
    \]
    \caption{Operational Semantic Rules for {\modediagram}}
    \label{tbl:inferenceRules}
\end{table}

The operational rules for {\modediagram} are given in Table~\ref{tbl:inferenceRules}. Here we adopt a big-step operational semantics for {\modediagram}, which means that we only observe the start and end points of a period in the current mode, while the state changes within a period are not recorded. This is reasonable since in practice  engineers usually monitor the states at the two ends of a period to decide if it works well. In the rules, we make use of an auxiliary function $\execute$ to represent the execution results for the mode in one period.
\[
\execute: \mathcal{CFG}(V) \times \mathcal{L} \times \State \times \mathbb{R}^+ \rightarrow \mathcal{L} \times \State
\]
It receives a flow graph, a program counter, an initial state and the time permitted to execute and returns the state and program counter after the given time is expired. Its detailed definition is left in the report \cite{modechart}. We now explain the operational rules:
\begin{itemize}
    \item[1.] \runa{enter}. When the system is at the beginning of a period, if the current mode $m$ has sub-modes, the system enters the initial sub-mode of $m$.

    \item[2.] \runa{detect}. When the system is at the beginning of a period, if the current mode $m$ is a leaf mode, the system updates its state by sampling from sensors. The function $\sampling$ represents the side-effect on variables during sensor detection. The period label $l$ is changed to be $\PExecute$, indicating that the system will then perform  computational tasks specified by the control flow graph of $m$.

    \item[3.] \runa{execute}. This rule describes the behaviors of executing CFG of the leaf mode $m$. \hide{The current state $\sigma$ is stored as the last element in the trace of states $\Sigma$. }The function $\execute$ is used to compute the new state $\sigma'$ from $\sigma$. The computation task may be finished in the current period and $pc' = \CFGExit$ holds or the task is not finished and the program counter points to some location in the control flow graph. The value of  the timestamp variable $ts$  in $\sigma'$ is equal to its value in state $\sigma$ plus the period of the mode $m$.

    \item[4.] \runa{continue}. This rule tells that when the computation task in leaf mode is not finished in a period, it will continue its task in the next period. In this case, the system is implicitly not allowed to switch to other modes from the current mode. When moving to  the next period, sensor detection is skipped.

    \item[5.] \runa{repeat}. This rule specifies the behavior of restarting the flow graph when the computation task is finished in a period. When it is at the end of a period and the system finishes executing the flow graph  ($pc = \CFGExit$), if there is no transition guard enabled, the system stays in the same mode and restarts the computation specified by the flow graph.

    \item[6.] \runa{switch}. This rule specifies the behavior of the mode transition. There exists a transition $t$, whose guard holds on the sequence of states $\Sigma$. And the priority of $t$ is higher than that of any other enabled transitions.

\end{itemize}

\section{The Property Specification Language}\label{sec:specification}

\begin{figure}[t]
  \centering
  \includegraphics[scale=0.5]{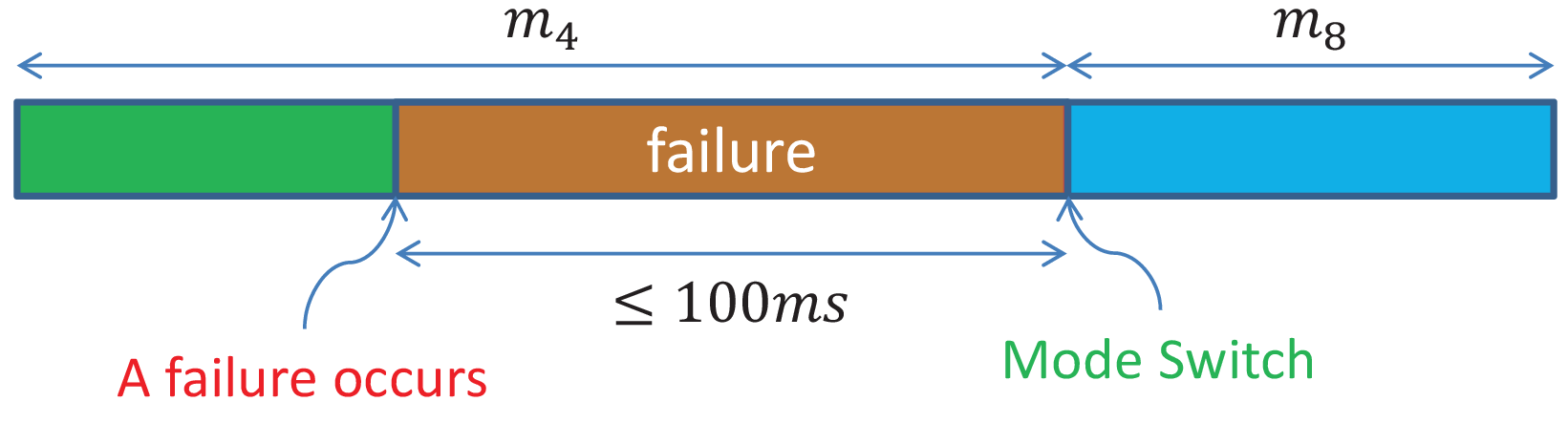}
  \caption{A Property about Failure}
  \label{fig:propertyExample}
\end{figure}

We adopt the Interval Temporal Logic (ITL)~\cite{ITLMaszkowskiM83} as the property specification language. The reason why we adopt the interval-based logic instead of state-based logics like LTL or CTL is that most of the properties the domain engineers care about are related to some duration of time. For instance, the engineers would like to check if the system specified by {\modediagram} can stay in a specific state for a continuous period of time  instead of just reaching this state. Another typical scenario illustrated in Fig.~\ref{fig:propertyExample} is that, ``\textit{when the system control is in mode $m_4$, if a failure occurs, it should switch to mode $m_8$ in 100 ms}''. The standard LTL formula $\Box(failure \wedge m_4 \Rightarrow \Diamond m_8)$ can be used to specify that ``\textit{when the system is in mode $m_4$, and a failure occurs, it should switch to mode $m_8$}''. But the real-time feature ``\textit{in 100 ms}'' is lost. Though the extended LTL or CTL may also describe the interval properties to some extent, it is more natural for the domain engineers to use interval-based logic since the intuitive chop operator ($^\frown$) is available in ITL.

An interval logic formula can be interpreted over a time interval \cite{IntervalCalculus} or over a ``state interval'' (a sequence of states)~\cite{ITLMaszkowskiM83} . \hide{For example,  that a formula $\phi ^\frown \psi$ holds in a time interval $[b, e]$ signifies that there exists $m$ ($b \leq m \leq e$) such that $\phi$ holds in interval $[b, m]$ and $\psi$ holds in $[m, e]$.  } As explained  later in this section, our proposed specification language will be interpreted in the latter way \cite{ITLMaszkowskiM83} except for a small modification on the interpretation of  the chop operator ($^\frown$).

\subsection{Syntax}

\begin{figure}[t]
  \centering
    $
        \begin{array}{crl}
              \textit{Terms}  &   \theta     \ \triangleq  & r \mid v \mid l \mid f(\theta_1, \ldots, \theta_n) \\
              \textit{Formulas} &   \phi, \psi \ \triangleq & \ \tvalue \mid \fvalue \mid p(\theta_1, \ldots, \theta_n)  \mid \neg \phi \mid \phi \wedge \psi \mid \phi ^\frown \psi \\
        \end{array}
    $
  \caption{The Syntax of ITL}
  \label{fig:syntaxITL}
\end{figure}

The syntax of the specification language is defined in Fig~\ref{fig:syntaxITL}, where
\begin{itemize}
    \item The set of terms $\theta$ contains real-value constants $r$, temporal variables $v$\hide{(whose values may depend on some time intervals)}, a special variable $l$, and  functions  $f(\theta_1, \ldots, \theta_n)$ (with $f$ being an $n$-arity function symbol   and $\theta_1,\ldots,\theta_n$ being terms). \hide{(1) Constant symbol $r$ denotes a real value. (2) Temporal symbol $v$ is used to represent the variable whose value depends on the given interval. (3) Function symbol $f(\theta_1, \ldots, \theta_n)$ means a n-arity real value function.}
    \item Formulae can be boolean constants  ($\tvalue$, $\fvalue$), predicates ($p(\theta_1, \ldots, \theta_n)$ with $p$, an $n$-arity predicate symbol), classical logic formulae (constructed using $\neg$, $\wedge$, etc), or interval logic formulae (constructed using $\CHOP$). \hide{(1) Constant symbols $\tvalue$, $\fvalue$ denote the boolean constants $\True$ and $\False$, respectively. (2) Predicate symbol $p(\theta_1, \ldots, \theta_n)$ means a n-arity predicate. (3) Boolean connections: negation($\neg$) and conjunction ($\wedge$) (4) Modal connection $\CHOP$ is a binary modality which \emph{chop}s an interval into two consecutive sub-intervals.} If the formula $\phi \CHOP \psi$ holds for an interval $\ell$, it means that the interval $\ell$ can be ``chopped'' into two sub-intervals, where $\phi$ holds for the first sub-interval and $\psi$  holds for  the second one.
\end{itemize}

As a kind of temporal logic, ITL also provides the $\Box$ and $\Diamond$ operators. They are defined as the abbreviations of $\CHOP$.

\[
\Diamond \phi \triangleq \tvalue \CHOP (\phi \CHOP \tvalue), \text{ for some sub-interval }, \ \
\Box \phi     \triangleq \neg \Diamond (\neg \phi), \text{ for all sub-intervals }
\]

By the specification language proposed here, we can describe the properties the domain engineers may desire. For instance, the following property describes the scenario shown in Fig.~\ref{fig:propertyExample}.

\[
    \Box(
    m_4 \wedge (\neg \failure \CHOP \failure) \CHOP \tvalue \Rightarrow
    m_4 \wedge (\neg \failure \CHOP (\failure \wedge l \le 100)) \CHOP m_8 \CHOP \tvalue
    )
\]


\begin{table}[t]
    \centering
    \begin{tabular}{l}
        $
        \begin{small}
            \begin{array}{l}
                \mathcal{I_T}(r, \Sigma) = r \\
                \mathcal{I_T}(l, \Sigma) = \begin{cases}
                                           \begin{array}{ll}
                                           \sigma_{n-1}(ts) - \sigma_{0}(ts) & \text{ if } \Sigma  = \sigma_0.\ldots.\sigma_{n{-}1} \\
                                           \infty                            & \text{ if } \mid \Sigma\mid = \infty \hide{\text{ is infinite }}
                                           \end{array}
                                           \end{cases} \\
                \mathcal{I_T}(v, \sigma_0 . \Sigma) = \sigma_0(v) \\
                \mathcal{I_T}(f(\theta_1, \ldots, \theta_n), \Sigma)  =  f\left(\mathcal{I_T}(\theta_1, \Sigma), \ldots, \mathcal{I_T}(\theta_n, \Sigma)\right) \\
            \end{array}
        \end{small}
        $ \\
        \\
        $
        \begin{small}
            \begin{array}{ccl}
                \mathcal{I_F}(p(\theta_1, \ldots, \theta_n), \Sigma) = \True & \text{ iff } &  p(\mathcal{I_T}(\theta_1, \Sigma), \ldots, \mathcal{I_T}(\theta_n, \Sigma)) \\
                 \mathcal{I_F}(\tvalue, \Sigma) = \True & \text{ iff } & \textit{always} \\
                \mathcal{I_F}(\fvalue, \Sigma) = \False & \text{ iff } & \textit{always} \\
                \mathcal{I_F}(\neg \phi, \Sigma) = \True & \text{ iff } & \mathcal{I_F}(\phi, \Sigma) = \False \\
                \mathcal{I_F}(\phi \wedge \psi, \Sigma) = \True & \text{ iff } &
                \mathcal{I_F}(\phi, \Sigma) = \True \text{ and } \mathcal{I_F}(\psi, \Sigma) = \True\\
                \mathcal{I_F}(\phi ^\frown \psi, \Sigma) = \True & \text{ iff } &
                \exists k  < \infty \cdot \ \Sigma = (\sigma_0 \ldots \sigma_k \cdot \Sigma') \wedge \\
                & & \hspace*{4mm} \mathcal{I_F}(\phi, \sigma_0\ldots\sigma_k) = \True \wedge \mathcal{I_F}(\psi, \Sigma') = \True \\
            \end{array}
        \end{small}
        $ \\
    \end{tabular}
  \caption{Interpretation of the Specification Language}\label{tbl:semanticsITL}
\end{table}

\subsection{Interpretation}

Terms/formulae in our property specification language are interpreted in the same way as in Maszkowski \cite{ITLMaszkowskiM83}, where
an interval is represented by  a finite or infinite sequence of states ($\Sigma = \sigma_0 \sigma_1 \ldots \sigma_{n-1} \ldots$), where $\sigma_i \in \State$. The interpretation is given by two functions (1) term interpretation :$\mathcal{I_T} \in  \Terms \times \Intv \mapsto \mathbb{R}$, and (2) formula interpretation function: $\mathcal{I_F} \in \Formulas \times \Intv \mapsto \{\True, \False\}$. Table~\ref{tbl:semanticsITL} defines these two functions, where $ts$  denotes the variable $\timestamp$. \hide{and a special temporal variable $l$ is introduced to denote the interval length. }The value of the variable $\timestamp$ increases with the elapse of the time. i.e.,  for any two states in the same interval $\sigma_i, \sigma_j$, if $i {<} j$, then $\sigma_i(ts) {<} \sigma_j(ts)$. Thus, we can compute the length of  time interval based on the difference of the two time stamps located in the first and last states respectively. The interpretation of a variable $v$ on $\Sigma$ is the evaluation of $v$ on the first state of $\Sigma$. Note that our chop operator requires that the first sub-interval of $\Sigma$ is restricted to be finite no matter whether the interval $\Sigma$ itself is finite or not.

\section{{\modediagram} Verification by Statistical Model Checking}\label{sec:statistical}

As a modelling \& verification framework for periodic control systems,  {\modediagram} supports the modelling of periodic behaviors, mode transition, and complex computations involving linear or non-linear mathematical formulae. Moreover, it also provides a property specification language to help the engineers capture requirements. In this section, we will show how to verify  that an {\modediagram} model satisfies  properties formalized in the specification language. There are two main obstacles to apply classic model checking techniques on {\modediagram}: (1) {\modediagram} models involve complex computations like non-linear mathematic formulae; (2) {\modediagram} models are open systems which need intensive interactions with outside.

Our proposed approach relies on the Statistical Model Checking(SMC)~\cite{SMCSenVA04,SMCYounes05,SMCTA2011, UppaalSMC}. SMC is a simulation-based technique that runs the system to generate traces, and then uses statistical theory to analyze the traces to obtain the verification estimation of the entire system. SMC usually deals with the following quantitative aspect of the system under verification~\cite{SMCYounes05}:
\begin{itemize}
    \item [] What is the probability that a random run of a system will satisfy the given property $\phi$?
\end{itemize}

Since the SMC technique verifies the target system with the probability estimation instead of the accurate analysis, it is very effective when being applied to open and non-linear systems. Since SMC depends on the generated traces of the system under verification, we shall  briefly describe how to simulate an {\modediagram} and then present an SMC algorithm for {\modediagram}.

\subsection{{\modediagram} Simulation}

 The {\modediagram} model captures a reactive system~\cite{ReactiveHarel}. The {\modediagram} model executes and interacts with its external environment in a \emph{control loop} in one period as follows: (1) Accept inputs via sensors from the environment. (2) Perform computational tasks. (3) Generate outputs to drive other components. The {\modediagram} simulation engine simulates the process of the control loop above.

Generally speaking, the simulation is implemented according to the inference rules defined in Table~\ref{tbl:inferenceRules}. However, the behaviors of an {\modediagram} model depends not only on the {\modediagram} itself, but also on the initial state and the external environment. When we simulate the {\modediagram} model, the initial values are randomly selected from a range specified by the control engineers from CAST. To make the simulation executable, we have to simulate the behaviors of the environment to make the {\modediagram} model to be closed with its environment. The environment simulator involving kinematical computations designed by the control engineers is combined with the {\modediagram} to simulate the physical environment the {\modediagram} model interacts with. In the end of each period, the guard of transitions is checked. The satisfaction of $\duration$ and $\after$ guards does not only depend on the current state, but also the past states. The simulator sets a counter for each $\duration$/$\after$ guard instead of recording the past states. As a {\modediagram} model is usually a non-terminating periodic system, the bound of periods is set during the process of simulation.

\subsection{SMC Algorithm}

\begin{figure}[t]
    \centering
    \begin{tabular}{rl}
    \textbf{input}  & $\MD$: the {\modediagram}, $\phi$: property, $B$: bound of periods \\
                    & $\delta$: confidence, $\epsilon$: approximation \\
    \textbf{output} & $p$: the probability that $\phi$ holds on an arbitrary run of $\MD$\\
    \textbf{begin}  & \\
      10 & $N := 4 * \frac{\log{\frac{1}{\delta}}}{\epsilon^2}$, $a := 0$ \\
      20 & \textbf{for} $i := 1$ \textbf{to} $N$ \textbf{do}\\
      30 & \hspace*{4mm} generate an initial state $s_0$ randomly \\
      40 & \hspace*{4mm} simulate the {\modediagram} from $s_0$ in $B$ periods to get the state trace $\Sigma$\\
      50 & \hspace*{4mm} \textbf{if} ($\mathcal{I_F}(\phi, \Sigma) = \True$) \textbf{then} $a := a + 1$\\
      60 & \textbf{end for} \\
      70 & \textbf{return} $\frac{a}{N}$ \\
      \textbf{end} &
    \end{tabular}
    \caption{Probability Estimation for \modediagram}\label{fig:smc}  \vspace*{-3mm}
\end{figure}
\underline{}

We apply the methodology in ~\cite{SMCYounes05} to estimate this probability that a random run of an {\modediagram} will satisfy the given property $\phi$ with a certain precision and certain level of confidence. The statistical model checking algorithm for {\modediagram} is illustrated in Fig.~\ref{fig:smc}. Since the run of the {\modediagram} usually is infinite, the users can set the length of the sequence by the number of periods based on the concrete application. This algorithm firstly computes the number $N$ of runs  based on the formula $N := 4 * \log(1/ \delta) / \epsilon^2$ which involves the confidence interval $[p - \delta, p + \delta]$ with the confidence level $1 - \epsilon$. Then the algorithm generates initial state (line 30) and gets a state trace $\Sigma$ by the inference rules defined in Table~\ref{tbl:inferenceRules} (line 40). The algorithm in line 50 decides whether $\phi$ holds on the constructed interval based on the interpretation for the specification language mentioned in Section~\ref{sec:specification}. If the interpretation is $\True$, the algorithm increases the number of traces on that property $\phi$ holds. Line 70 returns the probability for the satisfiability of $\phi$ on the {\modediagram}.

\section{Case Study and Experiments}\label{sec:implementation}

We have implemented the {\modediagram} modeling and verification framework and applied it onto several real periodic control systems.  And two of them (termed hereafter as {\SA} and {\SB}) are for spacecraft control developed by Chinese Academy of Space Technology (CAST), where the system {\SB} is an updated version of {\SA}. The  system {\SA} has been launched successfully while system {\SB} was under development when we conducted the study with CAST. The engineers in CAST would like to check if the system {\SB} is correctly updated. Both of the two systems have 17 modes (6 sub-modes). Mode $m4$ and mode $m7$ have sub-modes. Mode $m4$ has three sub-modes, $G0, G1$ and $G2$. Mode $m7$ also has three sub-mode, $S1$, $S5$ and $S10$. Fig.~\ref{fig:mcoverall} is a small portion of the {\modediagram} model for system {\SB}.

During the modeling phase by {\modediagram}, some existed ambiguities are detected in the requirement documents for both system {\SA} and {\SB}. For instance, the designers do not specify the priorities on the transitions between modes, while the engineers implement the priorities  based on their experiences. However, priority in {\modediagram} is a mandatory option, which could avoid the ambiguities for the transitions.

During the verification phase by the statistical model checking on {\modediagram}, two design defects in system {\SB} are uncovered by analyzing the verification results: (1)  A variable is not initialized properly. (2) A value from sensors is detected from the wrong hardware address. In the traditional developing process in CAST, these two defects may be revealed only after a prototype of the software is developed and then tested. Our approach can find such bugs in design phase and reduce the cost to fix defects. In the following, we explain the experiments on  the verification in more detail.



\subsection{Verified Properties}





We communicate with the engineers in CAST, summarize several properties the two classes of spacecrafts should obey, and present these properties in our specification language. A total of 12 properties are developed by the engineers and these properties are verified on the systems {\SA} and {\SB}. We only highlight three properties  because the verification results on these three properties are different for systems {\SA} and {\SB}, which reveal two defects.
\begin{property}
\textit{The system will eventually reach the stable state forever}
\[
tt ^\frown \Box(\sqrt{\omega_x^2 + \omega_y^2 + \omega_z^2} \leq 0.1 \wedge \sqrt{\dot{\omega_x}^2 + \dot{\omega_y}^2 + \dot{\omega_z}^2} \leq 0.01))
\]
where $\omega_x$, $\omega_x$ and $\omega_z$ are angles. $\dot{\omega_x}$, $\dot{\omega_x}$ and $\dot{\omega_z}$ are angle rates.
\end{property}

\begin{property}
\textit{The system starts from mode $m0$, and then it will finally switch to mode $m5$ or $m6$ or $m8$, and stay in one of these three mode forever}
\[
(\textsf{mode} = 0) ^{\frown} tt ^{\frown} \Box(\textsf{mode} = 5 \vee \textsf{mode = 6} \vee \textsf{mode} = 8)
\]
\end{property}

\begin{property}
\textit{Whenever the system switches to mode $m4$ and then leaves $m4$, during it stays in $m4$, it firstly stays in
sub-mode $G0$, and then it switches to sub-mode $G1$, and then $G2$.}
\[
    \begin{array}{c}
        \Box(\textsf{mode} \neq 4 ^{\frown} \textsf{mode} = 4 ^{\frown} \textsf{mode} \neq 4 \Rightarrow \textsf{mode} \neq 4 ^{\frown}\\
        \textsf{mode} = 4 \wedge (\textsf{gm} = 0 ^{\frown} \textsf{gm} = 1 ^{\frown} \textsf{gm} = 2)^{\frown} tt)
    \end{array}
\]
\end{property}

\subsection{Experimental Results}



For the parameters of the statistical model checking algorithm, we set the half length of confident interval to be $1\%$ ($\delta = 1\%$) and the error rate to be $5\%$ ($\epsilon = 5\%$). Based on this algorithm, the total $7369$ traces for each control system are required to be generated to compute the probabilities during the verification process. We also compute the probabilities under different period bounds based on the traces, from $500$ periods to $5000$ periods (the maximum bound 5000 is suggested from the engineers of CAST).

The average verification time for each property is about 5.4 hours (5000 periods). The maximum verification time for one property is 8.6 hours and the minimum verification time is 3.5 hours (5000 periods). The experiments are conducted on a workstation with eight-core Intel(R)Xeon(R) CPU E5345@2.33GHz, 16GB RAM and Windows Server 2003 operation system.

Twelve properties are all passed for system {\SA}, and nine of twelve properties are passed for system {\SB}.
Take the example for system {\SA}, the probabilities of properties
$P1$ and $P2$ increase from $0\%$ to $100\%$ with the periods bound
from $500$ to $5000$, which means it takes a couple of periods for
the system to reach a stable state. Given more periods, the
probability of stability is higher. The probabilities of $P2$ are
similar to those of $P1$. We communicate with the engineers and they
confirm that the experimental results $P1$ and $P2$ conform to the
control theory,

\begin{figure}[t]
    \centering
    \includegraphics[width=0.9\textwidth]{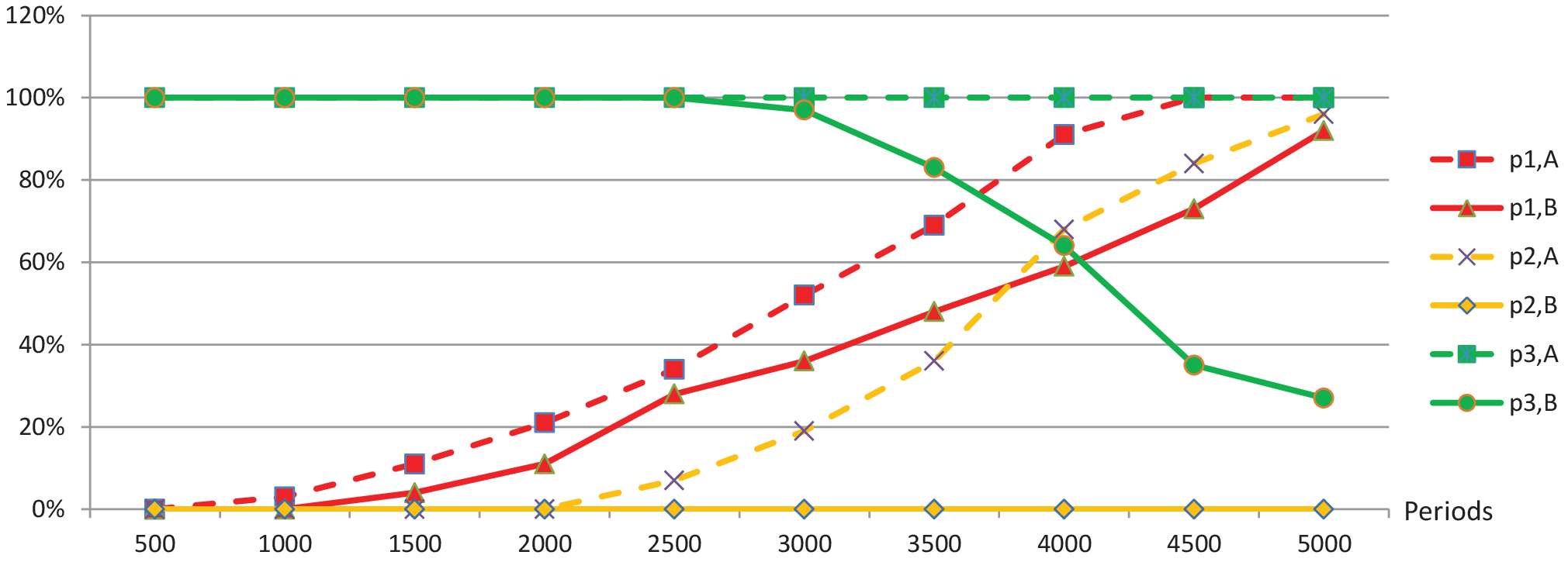}\\
    \caption{The Probabilities Curve}\label{fig:probability}\vspace*{-5mm}
\end{figure}

The experimental differences for systems {\SA} and {\SB} are for properties $P1, P2$ and $P3$, which are shown in Fig.~\ref{fig:probability}.
From this figure, we can see:
\begin{itemize}
    \item[1.] The probability of $P1$ in {\SB} also increases with the increasing of period. But it increases more slowly  in {\SB} than in {\SA}, and it only reaches $92\%$ most in the given period bound, which means control system {\SB} needs to take longer than what the control system {\SA} does to reach a stable state.
    \item[2.] For system {\SB}, the property $P2$ is not satisfied, for all the probabilities are $0\%$ in any period bound. But for system {\SA}, the probability of this property increases from $0\%$ to $100\%$.
    \item[3.] For system {\SB}, the probability of property $P3$ decreases from $100\%$ to $27\%$ with the increment of period bound, while for system  {\SA}, the probability of this property keeps $100\%$ in any period bound.
\end{itemize}

We communicate with the engineers from CAST to investigate why the verification results of system \emph{A} and \emph{B} are different. For property $P1$, we find that the difference  is caused by the unexpected value of a global variable $x$. The value of $x$ of system {\SB} is detected to be out of the specified range in some cases by analyzing the traces. In system {\SA}, before $x$ is used, its value is assigned by calling a preprocessing function. But in system {\SB}, the preprocessing function call is moved to a module which cannot be ensured to be called before using the variable $x$. As a result, $x$ is probably given a value outside the specified range, which leads to the slow arrival of the stable state in system {\SB}.

For property $P2$, by analyzing the traces of system {\SB} we find the system never enters modes $m5$, $m6$ and $m8$ from mode $m4$ and it stays in sub-mode $G1$ of mode $m4$ finally, which leaves the property $P2$  never satisfied.  The reason for leading to this problem is that the value of a variable $y$ is invalid. By the exploration of system {\SB}, we find that value of $y$ is still being assigned from the old hardware address while it should be obtained from the new one because the old bus is updated to a new one in system {\SB}. The engineers ignore the change in the design of system {\SB} when updating it from system {\SA}.


For property $P3$, we find the reason for that the probability of $P3$
decreases from $100\%$ to $27\%$ is that when the period bound is
short, the system does not switch to mode $4$ at all, which makes
the property $P3$ hold since its premise $\textsf{mode} \neq 4
^{\frown} \textsf{mode = 4} ^{\frown} \textsf{mode} \neq 4$ is
unsatisfied. With the increment of period bound, the system may
switch to mode $4$ in some cases but cannot reach the sub-mode $G2$,
so the probability of $P3$ decreases. The reason of the
unreachability of sub-mode $G2$ is the same with the one leading to
the unsatisfiability of the property $P2$.

\section{Related Work}\label{sec:discussion}

Our {\modediagram} can be broadly considered as a variant of Statecharts~\cite{statechart}, where a mode in  {\modediagram} is similar to a state in the Statecharts. However, we note the following distinctions: (1) In Statecharts, when a transition guard holds, the system immediately switches to the target state. But in {\modediagram}, mode switches are only allowed to be triggered at the end of a period.\hide{, since it is for the description of the periodic-driven system.} (2) In Statecharts, a transition guard is usually a boolean expression on the current(source) state; while in {\modediagram},  transition guards may involve past states via predicates like $\mathsf{during}$ and $\mathsf{after}$. (3) In Statecharts, all observations on the system are the states; while   {\modediagram} also concerns about the computation aspect of the system by means of the flow graphs provided in the leaf modes.

Giese et al.~\cite{rtsc} have proposed a semantics of real-time variant of Statecharts by introducing the Hierarchical Timed Automata. In another work~\cite{RealTimeUMLGiese} they have presented a compositional verification approach to the real time UML designs. A. K. Mok et al. have developed a kind of herarchical real-time chart named ``Modechart''~\cite{ModechartMok}. Compared with Giese et al.~\cite{rtsc}, parallel modes are supported in Modechart. Stateflow is the Statechart-like language used in the commercial software Matlab/Simulink~\cite{stateflow}. The Stateflow language enriches  Statecharts to allow it to support  flow-based and state-based computations together for specifying  discrete event systems. Our {\modediagram}  focuses more on  periodic control systems, which can be regarded as a specific type of  discrete event systems, and it provides the first class element \textit{period} to facilitate the precise modeling of periodic-driven systems. The transitions in Stateflow can be attached with a flowchart to describe  complicated computation,  the {\modediagram} specifies the flow graph for the computation in its leaf modes. While Stateflow focuses only on the modelling aspect of the systems,  the {\modediagram} integrates modelling and reasoning by providing a property specification language and a verification algorithm.

Some researchers  introduce \textit{operational mode}~\cite{OhMultiMode,SchmitzAE05Mode} during the modeling in hardware/software consynthesis. The operational mode is essentially a state in the automata, but it can be attached a flowchart for the description of the computation. It does not support the nested mode and period explicitly. However, it is actually an informal modeling notation because it allows to specify the system behaviors in the natural langauge. Our ModeChart is a lightweight formal notation for the modeling with its precise operational semantics.

Giotto is also a periodic-driven modeling language proposed by Henzinger et al.~\cite{giotto}. The main difference between Giotto and {\modediagram} is about the computation mechanism provided. The tasks in a mode can be performed in parallel in Giotto while  the details of the tasks are omitted and are moved to the implementation stage. The {\modediagram} supports the detailed description of the computation in their leaf modes since the design of it is targeted for  control systems which may involve rich algorithms. The {\modediagram} does not support the parallel computation explicitly at present since it could bring the nondeterminism at the design level. The emphasis of the Giotto is more the modeling and synthesis of  parallel tasks while the {\modediagram} is for the modeling and verification based on the proposed specification language.

Runtime Verification is a verification approach based on extracting information by executing the system and using the information to detect whether the observed behaviors violating the expected properties~\cite{RVHavelund,RVStolz}. The verification approach we apply in this paper is also a kind of runtime verification. But our methodology is the off-line analysis, while ~\cite{RVStolz} applies an on-line monitoring approach using Aspect-J. The reason to propose off-line analysis is that the cost to decide if an ITL formula is satisfiable on a given trace is huge, so information extraction and analysis are separated to two phases in our approach.

\section{Conclusion}\label{sec:conclusion}

In this paper, we propose the \underline{M}ode \underline{D}iagram \underline{M}odelling framework ({\modediagram}), a domain-specific formal visual modelling language  for periodic control systems. To support formal reasoning, {\modediagram} is equipped with a property specification language based on interval temporal logic and a statistical model checking algorithm. The property specification language allows engineers to precisely capture various properties they desire, while the verification algorithm allows them to reason about {\modediagram} models with respect to those properties. The viability and effectiveness of the proposed {\modediagram} framework have been demonstrated by a number of real life case studies (two of which have been presented in the paper), where defects of spacecraft control systems have been detected and uncovered in the early design stage.

\hide{, aiming to precisely capture various properties domain engineers may desire.  To verify the behaviors of the control systems specified by \modediagram, we develop a statistical model checking algorithm to check the properties against the \modediagram. The real life case studies show that the {\modediagram} is  useful for the control system engineering, and the experiments also show that real-world defects in the
system design can be uncovered by our approach. }

%
%
\bibliographystyle{abbrv}
\bibliography{main}

\end{document}